\documentclass[pre,aps,twocolumn,showpacs]{revtex4}
\usepackage{epsfig}
\include{graphics}
\begin{document}
\input{epsf}
\title{Shear-induced crystallization of a dense rapid granular flow: hydrodynamics
beyond the melting point?}
\author{Evgeniy Khain$^{1}$ and Baruch Meerson$^{2}$}
\affiliation{$^{1}$Department of Physics and Michigan Center for
Theoretical Physics, The University of Michigan, Ann Arbor,
Michigan 48109} \affiliation{$^{2}$Racah Institute of Physics,
Hebrew University of Jerusalem, Jerusalem 91904, Israel}

\begin{abstract}
We investigate shear-induced crystallization in a very dense flow of
mono-disperse inelastic hard spheres. We consider a steady plane
Couette flow under constant pressure and neglect gravity. We assume
that the granular density is greater than the melting point of the
equilibrium phase diagram of elastic hard spheres. We employ a
Navier-Stokes hydrodynamics with constitutive relations all of which
(except the shear viscosity) diverge at the crystal packing density,
while the shear viscosity diverges at a \textit{smaller} density.
The phase diagram of the steady flow is described by three
parameters: an effective Mach number, a scaled energy loss
parameter, and an integer number $m$: the number of
half-oscillations in a mechanical analogy that appears in this
problem. In a steady shear flow the viscous heating is balanced by
energy dissipation via inelastic collisions. This balance can have
different forms, producing either a uniform shear flow or a variety
of more complicated, nonlinear density, velocity and temperature
profiles. In particular, the model predicts a variety of multi-layer
two-phase steady shear flows with sharp interphase boundaries. Such
a flow may include a few zero-shear (solid-like) layers, each of
which moving as a whole, separated by fluid-like regions. As we are
dealing with a hard sphere model, the granulate is fluidized within
the ``solid" layers: the granular temperature is non-zero there, and
there is energy flow through the boundaries of the ``solid" layers.
A linear stability analysis of the uniform steady shear flow is
performed, and a plausible bifurcation diagram of the system, for a
fixed $m$, is suggested. The problem of selection of $m$ remains
open.
\end{abstract}
 \maketitle

\section{Introduction}

In spite of extensive experimental and theoretical studies of dense granular
flows, a theoretical description of these flows remains challenging
\cite{challenge}. Multi-particle contacts and friction, intrinsic in slow dense
flows, invalidate the kinetic theory \cite{Jenkins,hydrreview}. Furthermore,
even rapid dense flows (that is, flows dominated by binary collisions) present
significant difficulties for analysis. Dilute and moderately dense flows of
mono-disperse inelastic hard sphere fluids are describable, for not too high
inelasticity of collisions, by Navier-Stokes hydrodynamics \cite{hydrreview},
which can be derived in a systematic way from the kinetic theory: a one-particle
kinetic equation properly generalized to account for inelastic collisions
\cite{Jenkins,hydrreview}. Hydrodynamic equations, that is conservation laws for
the mass and momentum of the granulate, and a balance equation for the energy,
may still be valid, for a hard sphere fluid, at higher densities, after the
disorder-order transition occurs. However, the respective constitutive relations
are not derivable from a kinetic equation anymore.

In this work we attempt to address this difficulty and suggest a possible
hydrodynamic description of a sheared rapid granular flow that exhibits
crystallization. Shear-induced ordering in a dense granular medium has attracted
much recent attention. It was investigated experimentally (usually for slow
flows) by many groups \cite{Pouliquen,Gollub,Mueggenburg}. The crystallization
dynamics in inelastic hard sphere fluids has been also extensively studied in MD
simulations \cite{Polashenski,Alam}. In this work we will be dealing with an
idealized model of inelastic particle collisions characterized by a constant
coefficient of normal restitution $r$, and focus on a plane shear flow.

The present work is a next step in a series of recent attempts of
extending granular hydrodynamics of inelastic hard sphere fluids to
high densities \cite{Grossman,Bocquet,Meerson}. Grossman \textit{et
al.} \cite{Grossman} investigated a prototypical system of inelastic
hard disks at zero gravity, placed in a two-dimensional rectangular
box, one wall of which serving as a ``thermal" wall. Grossman
\textit{et al.} suggested an equation of state, granular heat
conductivity and inelastic energy loss rate which interpolated
between the dilute limit and the close vicinity of the hexagonal
close packing, where free-volume arguments are available. The
density profiles, obtained by solving the hydrostatic equations
numerically, were in good agreement with the results of MD
simulations \cite{Grossman}.

Meerson \textit{et al.} \cite{Meerson} considered a similar
two-dimensional granular system, but with gravity. The system was
driven from below by a ``thermal" base. Employing the constitutive
relations suggested by Grossman \textit{et al.} \cite{Grossman},
Meerson and coworkers found steady-state density and temperature
profiles and observed good agreement with MD simulations, including
the region of the ``levitating cluster", where the density is very
close to that of hexagonal close packing \cite{Meerson}.

Bocquet \textit{et al.} \cite{Bocquet} employed hydrodynamic
equations to model a granular shear flow where the density
approached the random close packing density. They also compared
their theory with experiment in a circular Couette flow
\cite{Bocquet}. As a shear flow was present, Bocquet and coworkers
had to specify, in addition to the rest of the constitutive
relations, the coefficient of shear viscosity. In experiment the
granular temperature had been found to decrease more slowly, with an
increase of the distance from the shear surface, than the velocity.
To account for this finding, Bocquet \textit{et al.} assumed that
the shear viscosity diverges more rapidly, at random close packing,
than the rest of the transport coefficients.

We employ in this work Navier-Stokes granular hydrodynamics for a
description of a steady crystallized shear flow of  an assembly of
mono-disperse inelastic hard spheres under constant pressure. Our
approach is similar to that taken by Bocquet \textit{et al.}
\cite{Bocquet}, with two important differences. First, we focus on
the ordered crystalline phase which ends at the crystal (hexagonal
or face-centered cubic) close packing density $\phi_{fcc}$, whereas
Bocquet \textit{et al.} considered the metastable disordered phase
which ends at the random close packing density $\phi_{rcp}$. Second,
we assume that the shear viscosity diverges at a \textit{smaller}
density than the rest of transport coefficients, see below. This
assumption brings about the possibility of a two-phase steady flow.
In general, a granular shear flow reaches a steady state when the
viscous heating makes up for the energy dissipation via inelastic
collisions. As we show here, this balance can be achieved in
different ways, producing either a uniform shear flow, or a variety
of flows with nonlinear density, velocity and temperature profiles.
Working in the range of densities beyond the melting point, we
determine the phase diagram of dense steady flows in terms of three
parameters: an effective Mach number, a scaled energy loss
parameter, and an integer number $m$: the number of
half-oscillations in a mechanical analogy that appears in this
problem. There are regions on this phase diagram where two or more
different steady flow solutions are possible for the same values of
the parameters. To get an additional insight, we perform a linear
stability analysis of the uniform steady shear flow. Based on these
results, we suggest a plausible bifurcation diagram of the system
which, in some region of the parameter space, describes bistability
and hysteresis.

The rest of the paper is organized as follows. In Sec. 2 we
introduce our shear-induced crystallization scenario and the
governing hydrodynamic equations and constitutive relations. Section
3 describes our model of a zero-gravity constant-pressure shear flow
and focuses on the analysis of a crystallized steady shear flow in
different regimes. Section 4 presents a linear stability analysis of
the uniform shear flow solution and suggests a plausible bifurcation
diagram of the system. Details of the linear stability analysis are
presented in the Appendix. Section 5 includes a brief discussion and
summary of our results.

\section{Pressure-density diagram and hydrodynamic equations}

\begin{figure}[ht]
\includegraphics[width=8.0 cm,clip=]{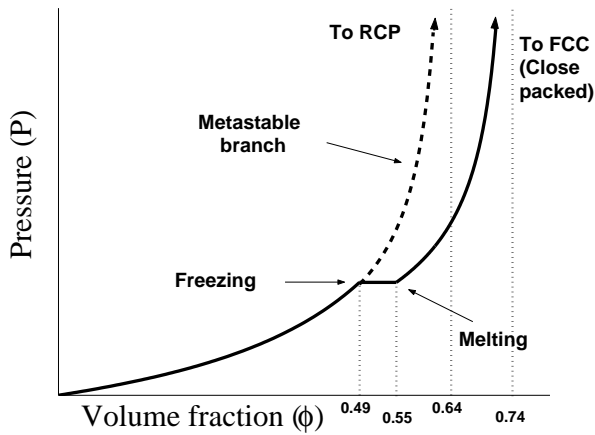}
\caption{Phase diagram of elastic hard sphere fluid, see text.}
\label{pressure}
\end{figure}

Figure~\ref{pressure} depicts, in the coordinates ``volume fraction
- pressure", the phase diagram of a homogeneous macroscopic system
of \textit{elastic} hard spheres \cite{Torquato}. The volume
fraction $\phi = (\pi / 6) d^3 n$ (where $n$ is the particle number
density, and $d$ is the particle diameter) changes from zero to
$\phi_{fcc}=\sqrt{2} \pi /6 \simeq 0.74$, the density of crystal
(either hexagonal, or face-centered cubic) close packing, the
densest possible packing of spheres in three dimensions. The phase
diagram includes four branches. The disordered, or gas/liquid branch
starts at $\phi = 0$ and continues until the freezing point that
occurs at $\phi \simeq 0.494$. Then this branch splits into two
branches. The horizontal (constant pressure) branch describes
coexistence of the disordered and ordered phases. It starts at the
freezing point and ends at the melting point at $\phi \simeq 0.545$.
At larger volume fractions the system is in the ordered crystalline
phase that ends at the density of the crystal close packing
$\phi_{fcc}$. The last branch is a {\it metastable} extension of the
gas/fluid branch. It starts at the freezing point and ends at random
close packing at $\phi_{rcp} \simeq 0.64$. Clearly, each branch is
described by a separate equation of state.

The phase diagram, presented in Fig.~\ref{pressure}, appears in the
context of a homogeneous system in either equilibrium, or a
metastable state. Granular systems are usually inhomogeneous and, no
less important, they are intrinsically far from equilibrium, due to
inelastic collisions between particles. We will deal, however, with
a small inelasticity of particle collisions and, in the spirit of
kinetic theory \cite{Jenkins,hydrreview}, assume that the system is
everywhere close to \textit{local} thermodynamic equilibrium, so the
phase diagram (and the constitutive relations presented below) are
valid locally. Let the fluid density be sufficiently large, so that
the volume fraction is everywhere above the melting point. There are
two possible phases here. One of them is the disordered phase (the
the metastable branch in Fig.~\ref{pressure}), the other one is the
crystallized phase. Each of the two phases, disordered and ordered,
can have inhomogeneous temperature and density profiles, without
violating the constancy of the pressure. Furthermore, domains of
disordered phase can in principle coexist with domains of the
ordered phases (again, without violating the constancy of the
pressure). Shear-induced crystallization apparently represents a
(non-equilibrium) phase transition, so that the metastable
disordered branch gives way, everywhere, to the stable ordered
branch.

This work  addresses very dense flows, where the volume fraction is
close to that of the crystalline close packing, $\phi_{fcc}-\phi \ll
\phi_{fcc}$. We assume that the ordered crystalline branch of the
phase diagram (see Fig.~\ref{pressure}) has already won in the
competition with the disordered branch. To describe a flow in this
phase, one needs hydrodynamic equations. These represent the mass
conservation
\begin{equation}
\frac{d n}{ d t} + n\, {\mathbf \nabla} \cdot {\mathbf v} =
0\,,\nonumber \label{cont}
\end{equation}
the momentum conservation
\begin{equation}
n \,\frac{ d{\mathbf v}}{d t} = {\mathbf \nabla} \cdot {\mathbf P}
+ \, n\, {\mathbf g}\,, \nonumber \label{momentum}
\end{equation}
and the energy balance
\begin{equation}
\frac{3}{2}\,n\, \frac{d T}{d t} =- {\mathbf \nabla} \cdot
{\mathbf Q} + {\mathbf P}\,: {\mathbf \nabla} {\mathbf v} -
\Gamma\,, \label{energy}
\end{equation}
where $\Gamma$ is the energy loss rate due to the inelasticity of
binary collisions. Here $n({\mathbf r},t)$ is the number density
of grains, $T({\mathbf r},t)$ is the granular temperature,
${\mathbf v}({\mathbf r},t)$ is the mean flow velocity, ${\mathbf
P}$ is the stress tensor, ${\mathbf g}$ is the gravity
acceleration, ${\mathbf Q}$ is the heat flux, and $d/dt =
\partial/\partial t + {\mathbf v} \cdot {\mathbf \nabla}$ is the
total derivative. In the following we will put the particle mass to
unity. The stress tensor ${\mathbf P}$ can be written as
\begin{equation}
{\mathbf P} =\left[ - p(n,T) + \mu(n,T)\,{\rm tr}\, ({\mathbf D}\,)
\right] {\mathbf I} + 2\eta(n,T)\, \hat{{\mathbf D}}\,,
\end{equation}
where
\begin{equation}
{\mathbf D} = (1/2)\left[{\mathbf \nabla v} + ({\mathbf \nabla
v})^{T}\right]
\end{equation}
is the rate of deformation tensor,
\begin{equation}
\hat{{\mathbf D}}={\mathbf D}-\frac{1}{3}\, {\rm tr}\, ({\mathbf
D}\,)\, {\mathbf I}
\end{equation}
is the deviatoric part of ${\mathbf D}$, and ${\mathbf I}$ is the
identity tensor. We assume that the heat flux ${\mathbf Q}$ is given
by the Fourier law
\begin{equation}
{\mathbf Q} = -\kappa(n,T)\,{\mathbf \nabla T}\,. \label{Fourier1}
\end{equation}
Similarly to dilute \cite{Fourier} and moderately dense
\cite{Lutsko2} rapid granular flows, there can be an additional term
in Eq.~(\ref{Fourier1}), proportional to the \textit{density}
gradient. This term must vanish as $r \to 1$, and it can be
neglected in the nearly elastic limit $q \ll 1$ that we are
interested in throughout this paper.

To make the formulation complete, one needs constitutive relations:
the equation of state $p=p(n,T)$ and the dependences of the
transport coefficients $\mu$, $\eta$ and $\kappa$, and of the energy
loss rate $\Gamma$ on $n$ and $T$. For small and moderate densities,
these relations can be derived from the Boltzmann or Enskog
equation, properly generalized to account for inelastic collisions
\cite{Jenkins,hydrreview}. For very large densities, that we are
interested in, one can use free volume arguments \cite{freevolume}.
The resulting equation of state is
\begin{equation}
p = p_1\,\frac{n_c^2\, T}{n_c-n}, \label{equationofstate}
\end{equation}
where $n_c=\sqrt{2}/d^3$ is the particle number density at close
packing, and $p_1$ is a numerical factor of order unity. Here and in
the following we will write $n_c$ instead of $n_{fcc}$
(face-centered cubic). The temperature dependence in Eq.
(\ref{equationofstate}) is exact. We assume that all transport
coefficients (except for the shear viscosity $\eta$, see below),
diverge like $(n_c-n)^{-1}$ near the close packing density
\cite{Grossman}. Therefore, one can write the bulk viscosity
coefficient $\mu(n,T)$, the thermal conductivity $\kappa(n,T)$, and
the energy loss rate $\Gamma (n,T)$ in the following form
\begin{eqnarray}
\mu = \mu_1\,\frac{n_c \, T^{1/2}}{(n_c-n)\,d^2}, \nonumber
\\
\kappa =\kappa_1\, \frac{n_c \, T^{1/2}}{(n_c-n)\,d^2},\nonumber
\\
\Gamma = \Gamma_1\, \left(1-r^2\right) \frac{n_c^2\,
T^{3/2}}{(n_c-n)\,d}\,. \label{constitutive}
\end{eqnarray}
Equations~(\ref{constitutive}) are valid in the limit $n_{c}-n \ll n_{c}$. The
temperature dependences are exact. The numerical factors $\mu_1$, $\kappa_1$,
and $\Gamma_1$ are of order unity and presently unknown; they can be found in MD
simulations. The same type of divergence of the transport coefficients (again,
except for the coefficient of shear viscosity $\eta$) was assumed in Ref.
\cite{Bocquet}, but in the vicinity of the \textit{random} close packing
density.

There is a recent evidence in the literature that the shear
viscosity coefficient $\eta$ of elastic hard disk fluid grows
faster with the density, at high densities, than the rest of the
transport coefficients \cite{Luding}. We believe that this
behavior remains qualitatively correct also for three dimensional
systems. Bocquet \textit{et al.} \cite{Bocquet} accounted for this
fact in their description of the plane shear flow near the random
close packing density. They proposed, by analogy with the behavior
of supercooled liquids above the glass transition, that the shear
viscosity coefficient diverges at random close packing density,
but with a larger exponent: $\eta \sim (n_{rcp} - n)^{-\beta}$,
$\beta>1$. In our model of a crystallized shear flow we suggest a
different approach. We will accommodate a recent finding of Luding
\textit{et al.} \cite{Luding} that the shear viscosity coefficient
diverges like $(n_{c}^{\ast} - n)^{-1}$ at a density
$n_{c}^{\ast}<n_c$:
\begin{equation}
\eta = \eta_1\,\frac{n_c \,T^{1/2}}{(n_c^{\ast}-n)\,d^2}\,,
\label{shearviscosity}
\end{equation}
where $\eta_1$ is a numerical factor of order unity, which is
presently unknown. Divergence of the shear viscosity implies that
the fluid is jammed on a \textit{macroscopic} length scale. While a
shear flow in a macroscopically jammed system is impossible, the
system of hard spheres may still have finite temperature, pressure,
heat conduction, and collisional energy loss.

There are arguments in the literature (see, \textit{e.g.}, Santos
\textit{et al.} \cite{Santos}) that steady sheared states of
granular fluids are intrinsically non-Newtonian (that is, require a
description beyond Navier-Stokes order). Why are such effects not
included in our description? The reason is that non-Newtonian
effects arise when one treats the inelasticity of particle
collisions in a non-perturbative way (postulating that the Boltzmann
equation, generalized to inelastic collisions, remains applicable
for finite inelasticity). In this paper we work in the limit of
nearly elastic particle collisions. In the first order in the
inelasticity one does not need to take into account any inelasticity
corrections to the transport coefficients (for example, to the shear
viscosity). The only place where the inelasticity enters in this
leading-order theory is the inelastic loss term $\Gamma$ in the
energy equation (\ref{energy}).

\section{Steady shear flow close to crystallization}
Now let us introduce the flow setting we will be dealing with in
the rest of the paper. We consider a plane Couette geometry. The
model system is infinite in the horizontal ($x$) direction and
driven by the upper wall $y=H$ that moves horizontally with
velocity $u_0$. The lower wall $y=0$ is at rest (of course, in the
absence of gravity the upper and lower walls are interchangeable).
The height $H$ of the layer in this setting is not fixed. Instead,
the upper wall is maintained at a constant pressure $P_0$, like in
a recent experiment by Gollub's group \cite{Gollub}. As already
mentioned above, we assume a very dense flow, so that the volume
fraction is close \textit{everywhere} to that of close packing:
$\phi_{fcc}-\phi \ll \phi_{fcc}$.

As we consider a steady horizontal motion, the number density $n$,
the granular temperature $T$ and the horizontal velocity $u$ depend
only on the vertical coordinate $y$. Then it follows from the
$y$-component of the momentum equation (\ref{momentum}) that the
pressure is constant throughout the system:
\begin{equation}
\label{puniform} p(y)=P_0\,.
\end{equation}
Now we write down the $x$-component of the momentum equation
(\ref{momentum}) and rewrite the energy balance equation
(\ref{energy}):
\begin{eqnarray}
\frac{d}{dy}\left(\eta \,\frac{du}{dy} \right) = 0, \nonumber
\\
\frac{d}{dy}\left(\kappa \frac{dT}{dy}\right) + \eta
\left(\frac{du}{dy} \right) ^ 2 - \Gamma (n,T) = 0\,,
\label{basiceq1}
\end{eqnarray}
where the constitutive relations are given by
Eqs.~(\ref{equationofstate})-(\ref{shearviscosity}).
Equations~(\ref{equationofstate})-(\ref{basiceq1}) must be
complemented by boundary conditions. We will assume rough walls and
no-slip boundary conditions for the horizontal velocity: $u(y=0)=0,
\, u(y=H) = u_0$. The problem of evaluation of the granular heat
flux at the rough walls was addressed by Chou \cite{Chou}. He
considered a model where inelastic spheres were driven by walls with
attached half-spheres (bumpy walls.) Extending an earlier treatment
by Richman \cite{Richman}, Chou calculated the heat flux and the
slip velocity at the boundaries. He showed that the heat flux at the
boundaries can be positive or negative, depending on whether the
``slip work" is larger or smaller than the energy loss due to
inelastic particle collisions with the walls. For some values of
parameters the total heat flux to the boundaries vanishes
\cite{Chou}. For simplicity, we will assume a vanishing heat flux
and therefore prescribe $dT/dy(y=0) = dT/dy(y=H) = 0$. The total
number of particles is conserved, which yields a normalization
condition for the density: $\int_0^H n(y)\,dy = N$, where $N$ is the
number of particles per unit area in the $xz$ plane.

Let us rescale the vertical coordinate $y$ by the ({\it a priori}
unknown) system height $H$, the horizontal velocity $u$ by the
upper plate velocity $u_0$, the density $n$ by the close-packing
density $n_c$, and the temperature $T$ by the ratio $P_0 / n_c$ of
the constant applied pressure and close-packing density. Rewriting
the equations in the scaled form, we obtain
\begin{eqnarray}
p_1\frac{T}{1-n}=1\,,\nonumber \\
\frac{d}{dy}\left(\frac{T^{1/2}}{1-\lambda-n} \,\frac{du}{dy}
\right) = 0\,, \nonumber
\\
\frac{d}{dy}\left(\frac{T^{1/2}}{1-n}\frac{dT}{dy}\right) +
\nonumber
\\
\frac{\eta_1}{\kappa_1}\frac{MT^{1/2}}{1-\lambda-n}
\left(\frac{du}{dy} \right)^2 -
\frac{\Gamma_1}{\kappa_1}\frac{2RT^{3/2}}{1-n} = 0\,,
\label{basiceq1a}
\end{eqnarray}
where $M = n_c\,u_0^2 /P_0$ is the square of the effective Mach
number of the flow, and $\lambda=1-n_c^{\ast}/n_c$ is a small
positive numerical factor. For simplicity, the three presently
unknown constants $p_1={\mathcal O}(1)$, $\eta_1/\kappa_1 ={\mathcal
O}(1)$ and $\Gamma_1/\kappa_1={\mathcal O}(1)$ are taken to be unity
in the following. As the system height is unknown {\it a priori},
the scaled quantity $R = (1-r^2)\,H^2/(2d^2)$ must be determined
from the solution of the problem. From the first of Eqs.
(\ref{basiceq1a}) we find a simple relation between the scaled
density and temperature: $T(y) = 1-n(y)$. Then, introducing a
convenient auxiliary variable $\epsilon(y) = [1-n(y)]^{1/2} \ll 1$,
we rewrite the remaining Eqs.~(\ref{basiceq1a}) as
\begin{equation}
\frac{du}{dy} = \sqrt{\frac{2c}{M}} \, \frac{\epsilon^2 -
\lambda}{\epsilon}\,, \label{basiceq3}
\end{equation}
\begin{equation}
\frac{d^2 \epsilon}{dy^2} + \left(c - R\right)\epsilon -
\frac{c\,\lambda}{\epsilon} = 0\,, \label{basiceq2}
\end{equation}
where $c$ is an unknown constant to be found from the solution of
the problem. The boundary and normalization conditions are
\begin{eqnarray}
u(y=0) = 0, \,\,\, u(y=1) = 1\,,  \nonumber
\\
\frac{d \epsilon}{dy}\left(y=0\right) =  \frac{d
\epsilon}{dy}\left(y=1\right) = 0\,, \nonumber
\\
\int_0^1 \left[1 - \epsilon(y)^2\right]\,dy = \frac{f}{R^{1/2}},
\label{bc}
\end{eqnarray}
where $f = (1-r^2)^{1/2}\,Nd^2/\,2$.  The steady flow equations
(\ref{basiceq3})-(\ref{bc}) include two (known) scaled parameters:
$M$ and $f$. There are five unknown parameters in the problem: $c$,
$R$ and three arbitrary constants which determine the solutions of
the ordinary differential equations (\ref{basiceq3}) and
(\ref{basiceq2}) for $u(y)$ and $\epsilon(y)$, respectively.
Correspondingly, there are five conditions (\ref{bc}) to determine
the unknown parameters. In the following, when solving the equations
numerically, we put $\lambda = 0.05$.

\begin{figure}
\includegraphics[width=7.0 cm,clip=]{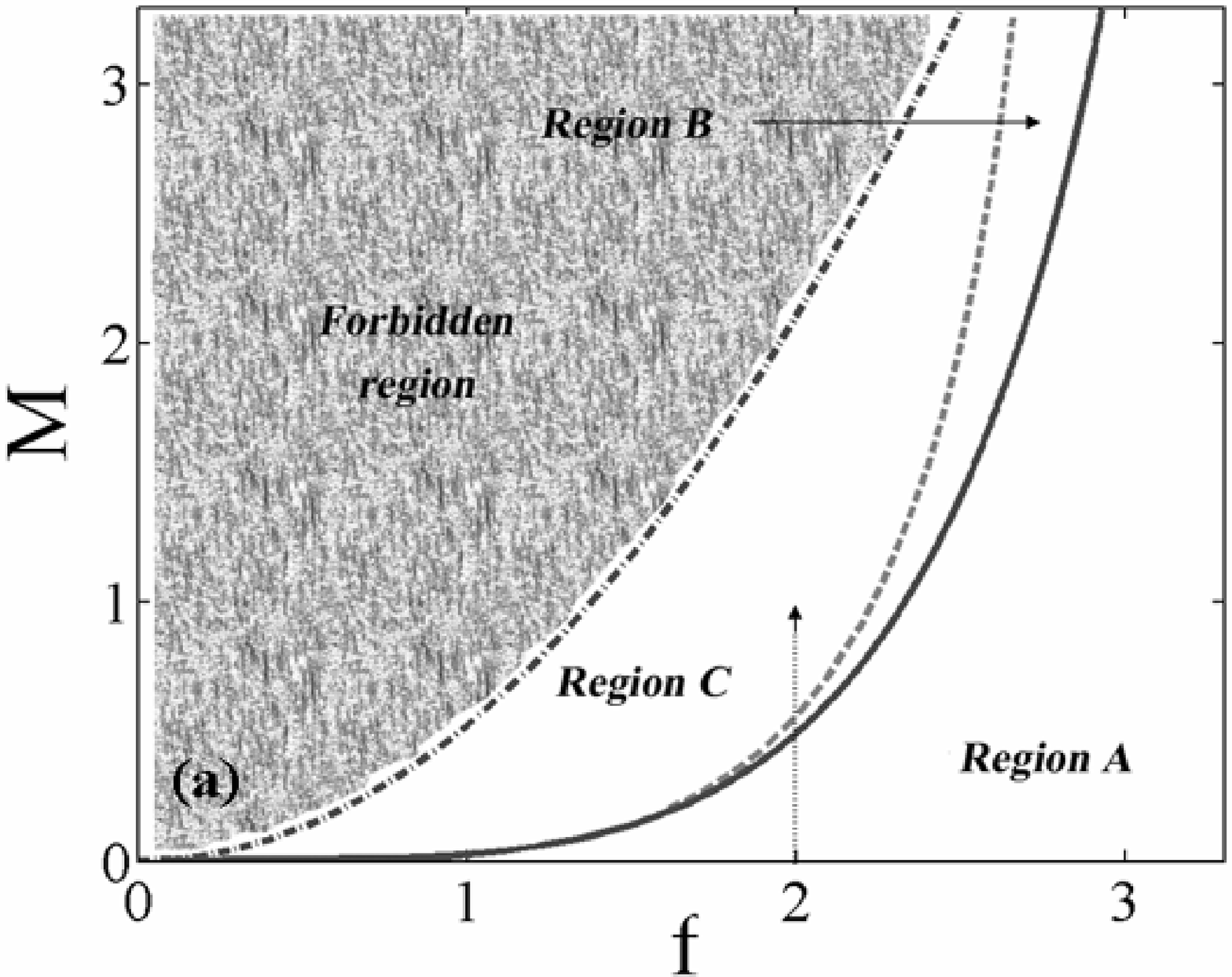}
\includegraphics[width=7.0 cm,clip=]{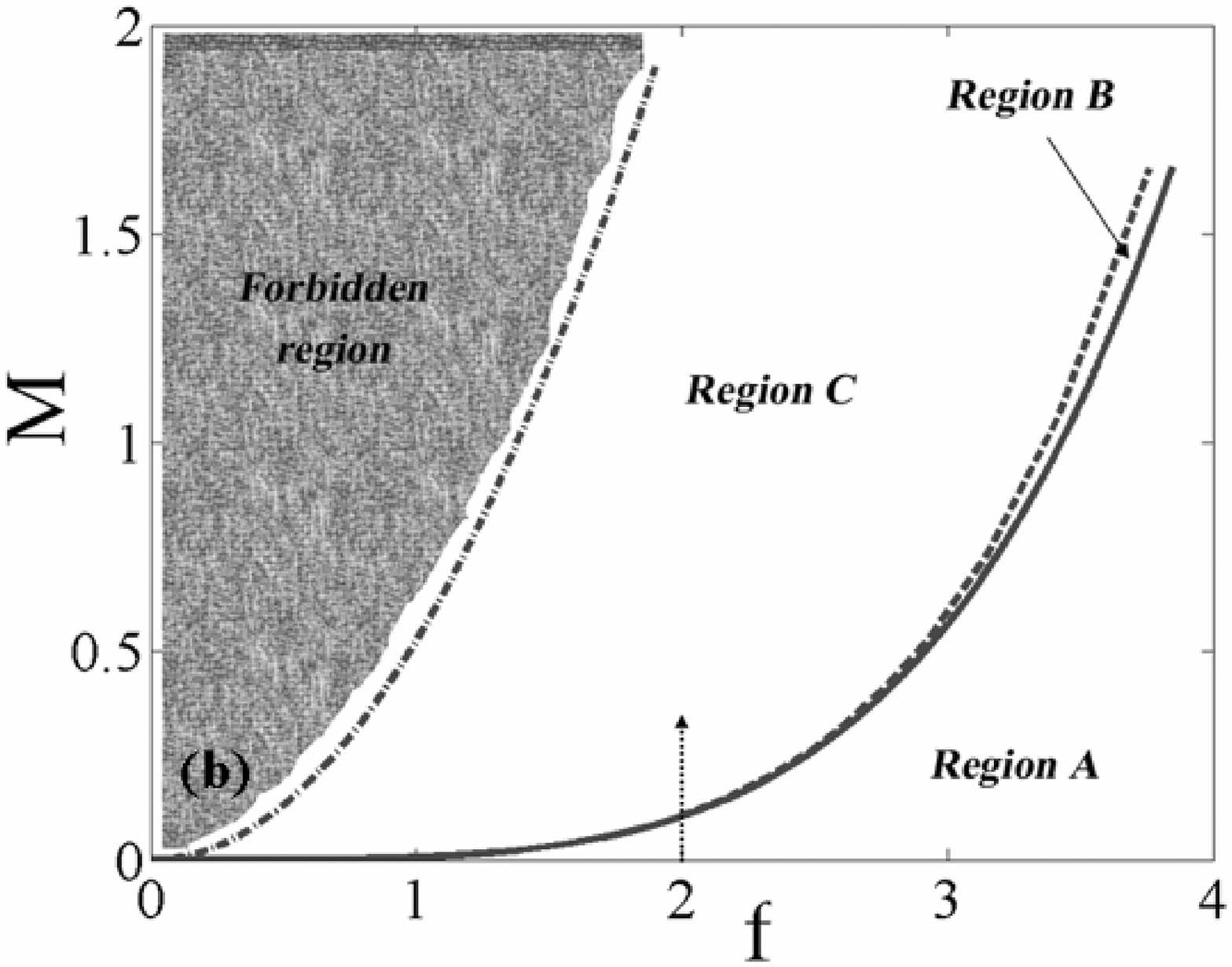}
\vspace{0.5cm} \caption{Hydrodynamic phase diagram of the steady
flow solutions for $m=1$ (the upper panel) and $m=2$ (the lower
panel) in terms of the scaled parameters $f$ and $M$. In the regions
A there are: a multi-layer flow and an (unstable) uniform flow. In
the regions B there are:  a multi-layer flow, a non-uniform flow and
an (unstable) uniform flow. In the regions C there is only a
(stable) uniform flow.  In the forbidden regions the granulate is
not dense enough for our theory to be valid. Plausible bifurcation
diagrams, corresponding to the dotted arrows $f = 2$, are shown, for
the cases of $m=1$ and $m=2$ in Fig.~\ref{bifurcation}.}
\label{diagram}
\end{figure}

Before we get into a detailed description of the steady flow
solutions of Eqs. (\ref{basiceq3})-(\ref{bc}), here is an overview
of the results. Strikingly, there is an infinite number of steady
flow solutions. They all can be parameterized  by \textit{three}
parameters: the scaled numbers $M$ and $f$ and an integer number
$m=1,2, \dots$. For a fixed $m$, there are three possible types of
solutions in different regions of the phase diagram $(M,f)$, see
Fig.~\ref{diagram}. The simplest solution is the uniform, or linear
shear flow that exists for any values of $M$ and $f$. Here the
velocity gradient, density and temperature are all constant. In
region $B$ (between the solid and dashed lines of
Fig.~\ref{diagram}), there is an additional one-phase solution: the
one with nonlinear profiles of density, temperature and velocity.
Finally, there is a multi-layer two-phase solution which exists in
regions $A$ and $B$, that is below the dashed line. The flow there
is organized in several distinct layers. In the ``solid-phase"
layers the density is larger than the critical density $n_c^{\ast}$
at which the shear viscosity diverges. Therefore, these layers move
as a whole with a velocity $u=const$, while the temperature and
density profiles are non-trivial. In the ``fluid phase" layers there
is a mean flow with non-trivial profiles of the velocity,
temperature and density. The density, temperature, heat flux and
velocity are continuous at the interface between the layers. This
means in particular, that there is no macroscopic flow in the bottom
layer of the granulate, and the inelastic energy losses there are
balanced by the conduction of heat from the (flowing) top layer.

Now let us consider these solutions in more detail. We start with
the simplest one: the uniform shear flow. Here $\epsilon$ is
independent of $y$, and Eq.~(\ref{basiceq2}) yields
$$\epsilon = \sqrt{\frac{c \lambda}{c - R}}\,.$$ Substituting this
value into Eq.~(\ref{basiceq3}) and integrating in $y$, we obtain
a linear velocity profile
\begin{equation}
u(y) = R\,\sqrt{\frac{2 \, \lambda}{M \, (c - R)}} \,\, y \,.
\label{velocity}
\end{equation}
Using the normalization condition and the condition $u(y=1)=1$, we
obtain
\begin{equation}
1 - \frac{c \lambda}{c - R} = \frac{f}{\sqrt{R}} \,\,\, \mbox{and}
\,\,\, c = R + \frac{2 \lambda R^2}{M}\,. \label{rc}
\end{equation}
Solving these two algebraic equations, we calculate $R$
\begin{eqnarray}
&&R=\frac{f^2}{2(1-\lambda)^2}\times \nonumber \\
&&\left[1+\frac{M(1-\lambda)}{f^2}+\left(1+\frac{2M(1-\lambda)}{f^2}\right)^{1/2}\right]
\label{R}
\end{eqnarray}
and obtain the scaled velocity profile and the constant value of
$\epsilon$:
\begin{eqnarray}
u_0 (y)&=& y,  \nonumber \\ \epsilon=\epsilon_\ast &=& \left[\lambda
+ \frac{(1-\lambda) k_1}{1 + k_1 + (1 +2k_1)^{1/2}}\right]^{1/2}\,,
\label{simplesolution}
\end{eqnarray}
where $k_1 \,= \,M\,(1-\lambda)\,f^{-2}$. Notice that
$\epsilon_\ast$ always satisfies the condition
$\epsilon>\lambda^{1/2}$ or, in the original variables,
$n<n_c^{\ast}$: the density of the uniform flow is smaller than the
critical density at which the shear viscosity diverges. Note also
that although the uniform shear flow solution
[Eq.~(\ref{simplesolution})] formally exists everywhere, the
assumption $\epsilon \ll 1$ demands $k_1 \ll 1$. Therefore, the
solution is valid for $M\,\ll f^2\,/(1-\lambda)$. This inequality
breaks down in the forbidden regions of the phase diagram, see Fig.
\ref{diagram}.

The nonlinear solution, which exists in region B, can be found
numerically. We used the following numerical procedure, realized in
Matlab. Let us denote $\epsilon_0 \equiv \epsilon(y=0)$. For fixed
$R$ and $f$, we first solve Eq.~(\ref{basiceq2}) by varying
parameters $\epsilon_0$ and $c$ and demanding $d\epsilon/dy(y=1)=0$
and the normalization condition. This procedure yields
$\epsilon(y)$.  Then we solve Eq.~(\ref{basiceq3}) and find the
velocity profile $u(y)$ and the corresponding value of parameter $M$
from the condition $u(y=1)=1$.

\begin{figure}[ht]
\includegraphics[width=6.5 cm,clip=]{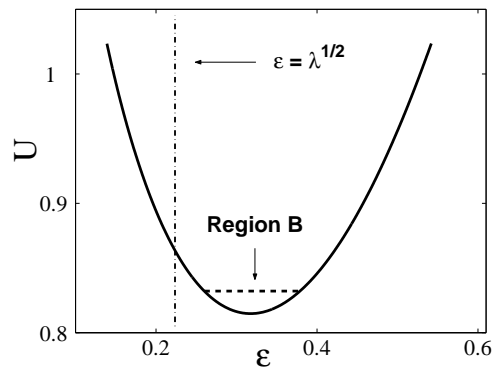}
\vspace{0.5cm} \caption{The mechanical analogy for
Eq.~(\ref{basiceq2}). Shown is a trajectory (the dashed line) of a
classical particle in the potential well $U(\epsilon)$ (the solid
line, see text) which corresponds to regions B in
Fig.~\ref{diagram}. See Figure~\ref{profilesB} for the resulting
spatial profiles, which correspond to a one-half of the oscillation
($m=1$) and a full oscillation ($m=2$, here the potential changes as
the parameter $M$ is different.) The oscillating solutions exist
only inside the potential well, and only for $\epsilon(y =
0)>\lambda^{1/2}$, when the density is everywhere below the critical
density at which the shear viscosity diverges. The parameters for
this figure are $M = 0.5158$ and $f = 2$.} \label{potential}
\end{figure}

\begin{figure}[ht]
\includegraphics[width=8.4 cm,clip=]{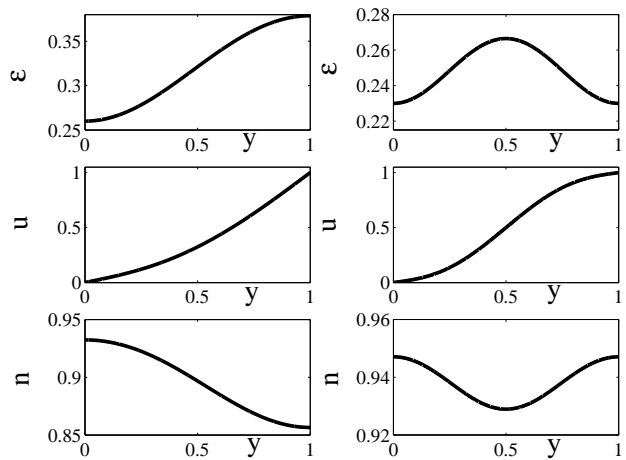}
\caption{The profiles of $\epsilon(y)=[1-n(y)]^{1/2}$ (the upper
row), the scaled velocity $u(y)$ (the middle row), and the scaled
density $n(y)$ (the bottom row), which correspond to region $B$ in
Figure~\ref{diagram}. The profiles in the left column (respectively,
in the right column) correspond to $m=1$, a one-half of a full
oscillation (respectively, to $m=2$, a full oscillation) in the
potential well, see Fig.~\ref{potential}. The parameters are $f=2$
and $M = 0.5158$ (the left column), and $f=2$ and $M = 0.1050$ (the
right column).} \label{profilesB}
\end{figure}

An important insight into the nature of this solution is provided by
a mechanical analogy following from Eq.~(\ref{basiceq2}). Let
$\epsilon$ be a ``coordinate", while the vertical coordinate $y$ be
``time". Then Eq.~(\ref{basiceq2}) describes a Newtonian particle
oscillating in the potential well $U(\epsilon) =
(1/2)(c-R)\,\epsilon^2 - c\lambda\ln\epsilon$. The boundary
conditions select a family of solutions. A typical solution includes
an integer number $m$ of halves of the oscillation period, and a
completion of these oscillations takes the system a unit ``time", $m
\,\tau/2 = 1$ [see Eq.~(\ref{period})], where $\tau$ is the
oscillation period. Figure~\ref{potential} shows an example of
particle in the potential well, while the resulting spatial profiles
of $\epsilon(y)$, $u(y)$ and $n(y)$ for a one-half of a full
oscillation ($m=1$), and for a full oscillation ($m=2$), are shown
in Fig.~\ref{profilesB}.

At a fixed $m$, this solution exists in region B of the phase
plane of parameters $f$ and $M$. This region is limited by two
curves $M=M(f)$ (see Fig.~\ref{diagram}). The first curve is
obtained as one approaches the bottom of the potential well. Here
the oscillation amplitude (that is, the density contrast in the
system) vanishes. Expanding the potential $U(\epsilon)$ near the
bottom of the potential well $\epsilon = [c \lambda/(c -
R)]^{1/2}$, we can calculate the oscillation period $\tau$:
\begin{equation}
\tau = \frac{2}{m} = \pi\,\left(\frac{2}{c-R}\right)^{1/2}\,.
\label{period}
\end{equation}
The bottom of the well corresponds to a uniform shear flow, but
Eq.~(\ref{period}) must be obeyed arbitrarily close to the bottom
of the well. Using Eq.~(\ref{period}) and the second equation of
Eqs.~(\ref{rc}), we obtain $R = (m\pi/2) \,(M/\lambda)^{1/2}$ and
$c = R+{m^2\pi^2}/2$. Substituting it into the first equation of
Eqs.~(\ref{rc}), we obtain the lower boundary of region B in terms
of $f$ and $M$:
\begin{equation}
f = \left(\frac{m\pi}{2}\right)^{1/2}
\left(\frac{M}{\lambda}\right)^{1/4}
\left[1-\lambda-\frac{(\lambda M)^{1/2}}{m\pi} \right]\,.
\label{lin}
\end{equation}
For $m=1$ and $m=2$ these curves are shown by the solid lines in
Fig.~\ref{diagram}. They determine the boundary between the
regions $A$ and $B$ for each $m$.

The second limiting curve (plotted by the dashed lines in
Fig.~\ref{diagram} for $m=1$ and $m=2$) is obtained numerically by
putting $\epsilon_0 = \lambda^{1/2}$ (see Fig.~\ref{potential}).
This curve determines the boundary between the regions $B$ and $C$.

\begin{figure}[ht]
\includegraphics[width=8.4 cm,clip=]{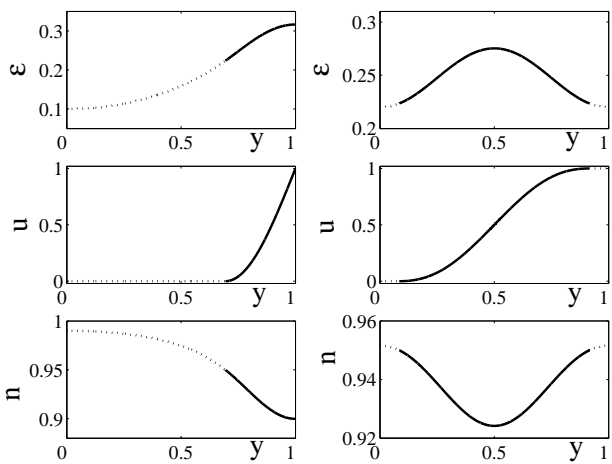}
\caption{The profiles of $\epsilon(y)$ (the upper row), the velocity
$u(y)$ (the middle row) and the density $n(y)$ (the bottom row) for
the two-phase solution, for $m=1$ (the left column) and $m=2$ (the
right column). The solutions for the fluid (solid) layers are shown
by the solid (dotted) lines. The parameters are $f = 2$ and $M =
0.0492$ (the left column), and $f = 2$ and $M = 0.10148$ (the right
column.)} \label{twophase}
\end{figure}

Now let us consider the most interesting multi-layer two-phase
family of solutions, parameterized by the same number $m\geq 1$ that
now can be associated with the number of ``solid" layers. Due to
continuity of the velocity field, there is no macroscopic flow in
the bottom layer of the granulate for any $m$, while the inelastic
energy losses there are balanced by the heat conduction from the
fluid-like top layer. A typical flow here consists of $m$ zero-shear
(solid-like) layers, each of which moving as a whole, separated by
fluid-like regions. As we are dealing with a hard-sphere model, the
granulate is fluidized within the ``solid" layers: the granular
temperature is non-zero, and there is energy flow through the
boundaries of the layers. Let us first consider the $m=1$ solution.
Here the only solid-like layer is at the bottom, and it is at rest.
Putting $c=0$ in Eqs.~(\ref{basiceq2}) and (\ref{basiceq3}) we
obtain $\epsilon(y) = \epsilon_0 \cosh(R^{1/2}y)$. The height of the
bottom layer is determined from the condition $\epsilon(h) =
\lambda^{1/2}$. At the interface between the two layers we demand
continuity of the density (and, therefore, of the temperature), of
the heat flux, and of the velocity. Similarly to the procedure
performed in region $B$, we solve the problem numerically by
shooting in two parameters $\epsilon_0$ and $c$ (for the top layer)
for fixed values of $R$ and $f$, and then calculate the profiles and
the respective value of the parameter $M$. Typical profiles
$\epsilon(y)$, $u(y)$ and $n(y)$ for $m=1$ are shown in
Fig.~\ref{twophase}, the left column. A multi-layer solution $m>1$
is obtained in a similar way, by demanding that the velocity in any
solid layer is constant, while the density, temperature, heat flux
and velocity are continuous at all interfaces between the layers.
Typical profiles of $\epsilon(y)$, $u(y)$ and $n(y)$ for $m=2$ are
shown in Fig.~\ref{twophase}, right panel.

At this point let us return, for a moment, to the case where the
viscosity singularity is assumed to be at the same density $n_c$ as
the rest of transport coefficient. In this case $\lambda = 0$, and
the only possible solution is a uniform shear flow. Indeed, as
$\epsilon$ must be positive, the only acceptable solution of
Eq.~(\ref{basiceq2}) with $\lambda=0$, which obeys the no-flux
boundary conditions, is $\epsilon=const$ and $c = R$. Then the
normalization condition [the last of Eqs.~(\ref{bc})] yields
$\epsilon^2=1-f/R^{1/2}$. One can obtain this solution directly by
putting $\lambda = 0$ in Eq.~(\ref{simplesolution}). Therefore, the
assumption of a nonzero $\lambda$ [the specific form of viscosity
divergence, Eq.~(\ref{shearviscosity})], is crucial for the
existence of non-trivial solutions.

One can see that, at fixed $m$, two different kinds of steady flow
solutions exist in region A, for the same values of parameters $M$
and $f$. Furthermore, three different kinds of steady flow solutions
exist in region B, again for the same values of parameters $M$ and
$f$ (see Fig.~\ref{diagram}). What is the selection rule for these
solutions? We give a partial answer to this question in the next
section by performing a linear stability analysis of the uniform
dense shear flow. Then we suggest plausible bifurcation diagrams of
the system for different $m$.

\section{Linear stability and bifurcations}

\begin{figure}[ht]
\includegraphics[width=7.0 cm,clip=]{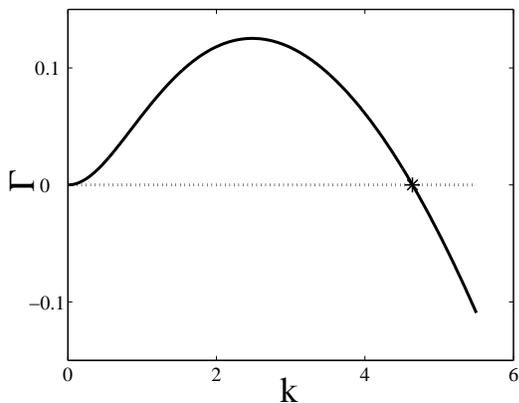}
\vspace{0.5cm} \caption{The scaled growth rate $\Gamma$,
calculated from Eq.~(\ref{quadr}), versus the scaled wave number
$k$. The long-wavelength modes with $k<k_{\ast}$ (the value of
$k_{\ast}$ is indicated by the asterisk) are unstable. The
parameters $f = 2$, $M = 0.2$, $\lambda = 0.05$, and $L = 100$
correspond to region $A$ in Fig.~\ref{diagram}a and to region C in
Fig.~\ref{diagram}b.} \label{dispersion}
\end{figure}

Our linear stability analysis of the uniform flow deals with the
full set of (time-dependent) hydrodynamic equations
(\ref{cont})-(\ref{energy}) and constitutive relations
(\ref{equationofstate})-(\ref{shearviscosity}). The details of
linear stability analysis are shown in Appendix. Adding a small
perturbations to the uniform shear flow and linearizing the
equations, we finally arrive at a quadratic characteristic
equation for the growth/damping rate $\Gamma$ as a function of
parameters and the wave number $k$ [Eq.~(\ref{quadr}), see
Appendix.] The two roots of this equation are real and correspond
to two different collective modes of the system. One of them
always decays. The other one is a purely growing mode for
sufficiently small $k$, that is for long-wavelength perturbations.
At small $k$ we obtain:
\begin{equation}
\Gamma = \frac{2\,R^2\,\lambda\,k^2}{M\,L\,\epsilon_\ast\,(M +
f\,R^{1/2})}>0\,, \label{Gamma}
\end{equation}
where $L=(2\,M)^{1/2}\,H/d$. For sufficiently large $k$, $\Gamma$
is negative: the heat conduction suppresses the instability. The
critical wave number for the instability is $k_\ast =
2R\lambda^{1/2}/M^{1/2}$. The dependence of the scaled growth rate
$\Gamma$ on the scaled wave number $k$ is shown in
Fig.~\ref{dispersion}.

Using Eq.~(\ref{R}), we obtain
\begin{eqnarray}
&&k_\ast =\frac{f^2\,\lambda^{1/2}}{2(1-\lambda)^2\,M^{1/2}}
\times
\nonumber \\
&&\left[1+\frac{M(1-\lambda)}{f^2}+\left(1+\frac{2M(1-\lambda)}{f^2}\right)^{1/2}\right].
\label{kstar}
\end{eqnarray}
Notice that the time scale separation, employed here for the
reduction of the order of the dispersion equation (see Appendix),
breaks down when $\epsilon_{\ast}$ approaches $\lambda^{1/2}$. As
one can see from Eq. (\ref{simplesolution}), this happens, for a
fixed $f$, when $M$ becomes sufficiently small. Importantly, the
final results (\ref{Gamma}) and (\ref{kstar}) remain valid in the
general case, as we obtained from the full, unreduced fourth-order
dispersion equation for $\Gamma(k)$.

\begin{figure}[ht]
\includegraphics[width=6.5 cm,clip=]{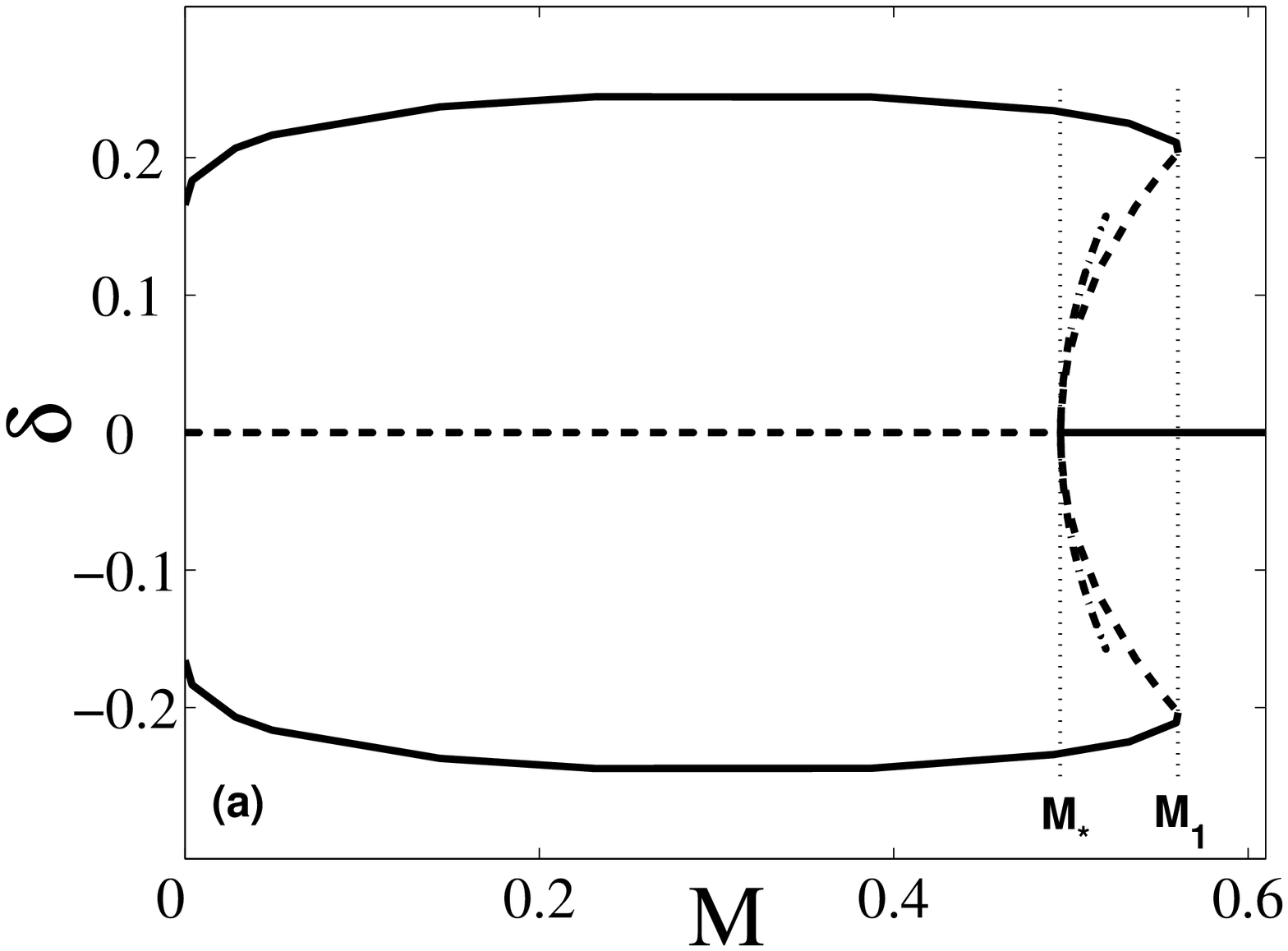}
\includegraphics[width=6.5 cm,clip=]{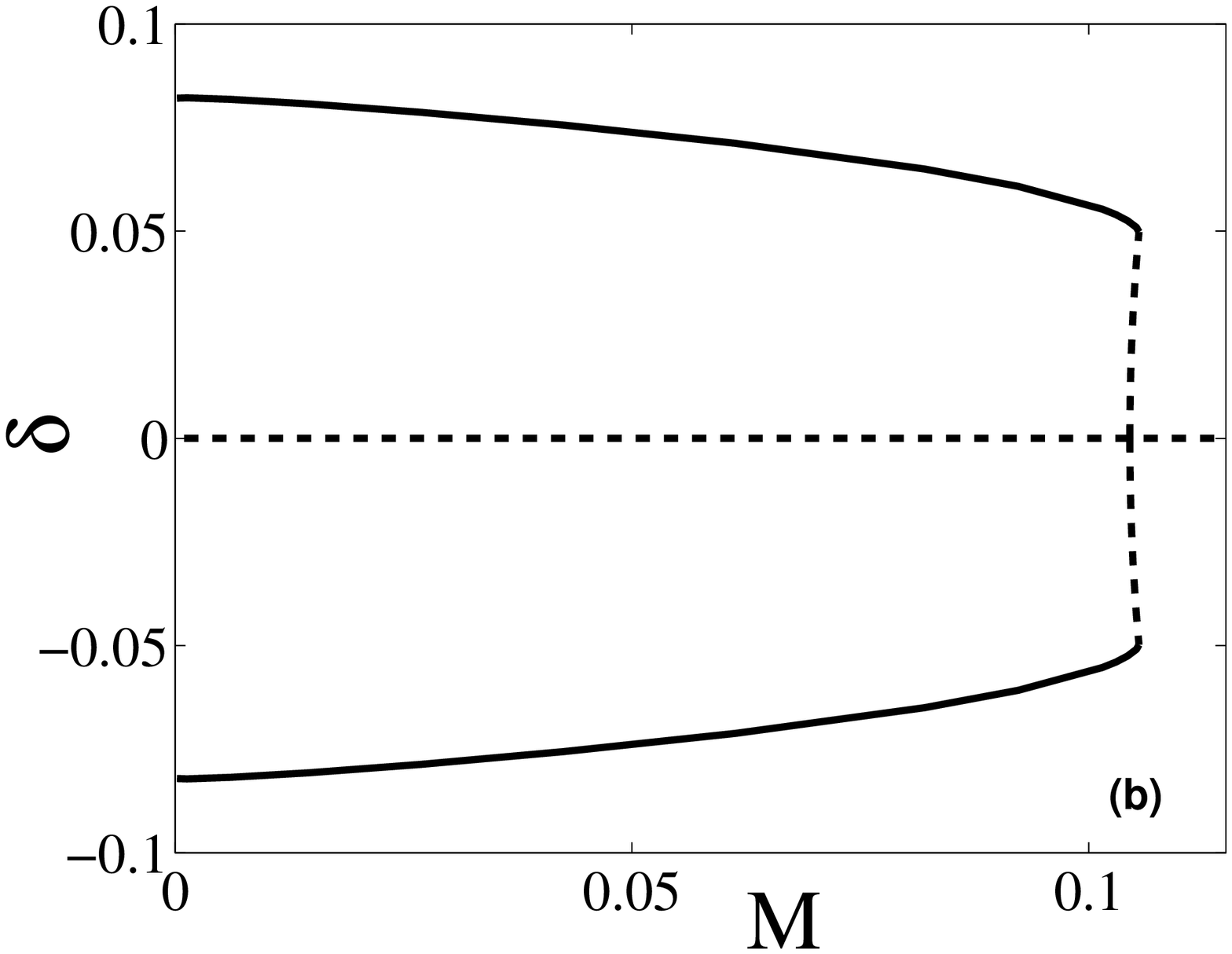}
\caption{A conjectured bifurcation diagrams of the steady flow for
$f = 2$ and $\lambda = 0.05$. Shown is the density contrast $\delta$
versus $M$ for different steady flow solutions for $m=1$ (the upper
panel) and $m=2$ (the lower panel). The stable and unstable branches
are indicated by the solid and dashed lines, respectively. We proved
that the uniform flow ($\delta=0$) is unstable at $M<M_{\ast}$. The
dash-dotted line in the upper panel is the asymptote of the branch
bifurcating from the point $M=M_{\ast}$: $\delta=0.9856\dots (M -
M_\ast)^{1/2}$.} \label{bifurcation}
\end{figure}

The wave number $k$ of the perturbation is quantized by the boundary
conditions. Indeed, the velocity perturbations must vanish at the
upper and lower plates:
$u_1(y=0,t)=u_1(y=1,t)=v_1(y=0,t)=v_1(y=1,t)=0$. These conditions
yield a discrete spectrum of wave numbers: $k=\pi m$, where
$m=1,2,3, \dots$. Therefore, the instability threshold is
$k_\ast=\pi$. This equation determines a curve $f(M)$ on the $(f,M)$
plane. One can see that this curve coincides with the curve $f(M)$
determined by Eq.~(\ref{lin}) for $m=1$: the borderline between
regions $A$ and $B$ in Fig.~\ref{diagram}. Similarly, the threshold
for $m=2$ perturbation $k_\ast=2\pi$ coincides with the curve $f(M)$
determined by Eq.~(\ref{lin}) for $m=2$, and so on. This is not
entirely surprising: as the instability is aperiodic, its threshold
is provided by the marginal stability condition.  It is convenient
to characterize each steady flow solution by the maximum density
contrast it predicts. In terms of the auxiliary variable $\epsilon$,
we can define $\delta$ as the maximum change in $\epsilon$
throughout the system [for $m=1$ this gives $\delta \equiv
\epsilon(y=1)-\epsilon(y=0)$.] We can calculate $\delta$ as a
function of parameters $f$ and $M$ for each steady flow solution.
Figure~\ref{bifurcation} shows all the resulting branches of the
solution (for $m=1$ and $m=2$) for a fixed $f$, while $M$ serves as
a control parameter. Furthermore, there is an infinite number of
bifurcation diagrams for multi-layer solutions (for all integers
$m>1$), which can be computed in a similar way.

In order to determine the bifurcation diagram of the system at a
fixed $m$, one needs to perform a linear stability analysis of each
branch of the solution. Unfortunately, such an analysis is quite
cumbersome for the solutions with $\delta \neq 0$. It is natural to
assume that bifurcation, which occurs in the vicinity of the point
$M=M_{\ast}$, at which the uniform shear flow (the one with
$\delta=0$) changes from unstable [at $M<M_{\ast}(f)$] to stable [at
$M>M_{\ast}(f)$], is an inverse pitchfork bifurcation. Based on this
assumption, we obtain a simple bifurcation diagram, presented in
Fig.~\ref{bifurcation}. The stable solutions are indicated by solid
lines, the unstable solutions by dashed lines. Let us consider the
case of $m=1$ and follow the uniform shear flow solution (for which
$\delta = 0$) as $M$ increases, starting from a small value, at
fixed $f$. We pass from region $A$ to region $B$ of the hydrodynamic
phase diagram (see the dotted arrow in Fig.~\ref{diagram}) via  an
inverse pitchfork bifurcation. Here the uniform shear flow solution
becomes stable, while the bifurcating ``second solution" (the one
with nonlinear density and velocity profiles) must be unstable.
Exploiting the mechanical analogy (see Fig.~\ref{potential}), we can
find the asymptote of the unstable branch in the close vicinity of
the bifurcation point: $\delta = A(f) (M - M_\ast)^{1/2}$. Now let
us move along the unstable branch $\delta \neq 0$ and increase $M$.
At some critical value of $M$, which depends on $f$, we reach the
border between regions $B$ and $C$. Here the unstable branch ends,
as for larger $M$ (region $C$) the only possible solution is the
uniform shear flow. There is, however, another solution which exists
at smaller $M$. Figure~\ref{bifurcation} shows this \textit{stable}
branch which corresponds to the two-phase solution in regions $A$
and $B$ (see Fig.~\ref{diagram}). This simple scenario predicts
bistability at sufficiently small values of the parameter $M$, and a
hysteresis on the interval $[M_{\ast}(f), M_1(f)]$, see
Fig.~\ref{bifurcation}. Note that the parameter $L$ does not affect
the bifurcation diagram, it only sets the time scales of transient
motions in the system.

\section{Summary and discussion}

We have considered here some aspects of shear-induced
crystallization in a dense but rapid mono-disperse granular shear
flow. We focused on a steady crystallized flow under a constant
pressure and zero gravity. Assuming very high densities, $n_c-n \ll
n_c$, we employed a version of the Navier-Stokes hydrodynamics for
inelastic hard sphere fluid with \textit{ad hoc} constitutive
relations based on the free volume argument. In contrast to earlier
works on rapid granular shear flow, we assumed that the shear
viscosity coefficient $\eta$ diverges at a density smaller than the
close packing density $n_c$, while the rest of the constitutive
relations diverge at $n=n_c$. We have determined the phase diagram
of the steady flow in terms of three parameters: the effective Mach
number, the scaled inelastic energy loss parameter, and an integer
number $m$. In a steady flow the viscous heating of the granulate is
balanced by energy dissipation through inelastic collisions. This
balance is achieved, in different parts of the phase diagram, in
different ways, producing either a uniform shear flow (with constant
velocity gradient, density and temperature), or a flow with
nonlinear velocity, density and temperature profiles. In some
regions in the phase diagram two or three different steady flow
solutions are possible for the same values of the parameters. We
performed a linear stability analysis of the uniform flow, and
suggested a plausible bifurcation diagram of the flow at a fixed
$m$, which predicts bi-stability and hysteresis. We are unable as
yet to find a selection principle that would prefer certain steady
state solutions out of a multitude of solutions at different $m$.
This non-trivial selection problem should be addressed in the future
work.

One of the predictions of this work is the existence (and, we
conjecture, stability) of two-phase solutions. The simplest solution
of this type consists of a zero-shear (solid-like) layer at the
bottom and a flowing top layer. Though there is no mean flow in the
bottom layer, the particles there undergo ``thermal" motion, and the
granular temperature and pressure are non-zero. As a result, there
is energy transfer through the bottom layer. There also exist
two-phase multi-layer solutions, where solid-like layers, each of
which moving as a whole, are separated by fluid-like regions with
nonlinear velocity, density and temperature profiles. The existence
of these solutions is a direct consequence of our assumption that
the coefficient of shear viscosity $\eta$ in
Eq.~(\ref{shearviscosity}) diverges at a density which is smaller
than the close packing density $n_{c}$.

A comparison of our results with those of Alam and coworkers is in
order now. Alam {\it et al.} \cite{Alam1} investigated a plane
Couette shear flow of inelastic hard sphere fluid, in a system with
a fixed height, in a wide interval of densities: from the dilute
limit to the random close packing density. They employed
constitutive relations, all of which (including the viscosity)
diverge at the random close packing density $n_r$. They found
instability of the uniform shear flow when the inelasticity of the
particle collisions becomes large enough or, alternatively, when the
(fixed) system height exceeds a critical value for a fixed
inelasticity. The uniform shear flow instability considered in our
work is quite different: it \textit{requires} viscosity divergence
at a smaller density than the rest of constitutive relations.
Indeed, Eqs.~(\ref{lin}) and (\ref{kstar}) show that this
instability disappears when $\lambda$ goes to zero. Although the
parameter $\lambda$ was identically zero in Ref. \cite{Alam1},
instability was observed. Where does the difference come from? To
remind the reader, the constitutive relations that we used assume a
very dense system. Respectively, our results are valid only at
leading order in the parameter $(n_c-n)/n_c$. We checked that, if
one takes into account only the leading order terms in $(n_r-n)/n_r$
in the equations of Alam \textit{et al.} \cite{Alam1}, the
instability disappears.

Its worth mentioning that the steady flow equations (\ref{basiceq3})-(\ref{bc})
would not change if the crystal close packing density $n_c$ were replaced by the
random close packing density $n_r$. Such a formulation would follow from the
assumption that the shear viscosity diverges at a density smaller than $n_r$. It
would predict a variety of steady state solutions on the metastable branch, in
the vicinity of the random close packing. We hope that the basic assumptions of
our model (including the specific form of viscosity divergence) and its
non-trivial predictions will be tested in MD simulations and in experiment on
dense but {\it rapid} granular flow. Our work focused on a dense but rapid
granular flow, assuming that the granulate is fully fluidized. In experiment it
should be easier to achieve this regime when the shear is very large, so that
the effective Mach number is of order unity (see Fig.~\ref{bifurcation}.) Most
experiments with dense granular flows are performed with slow flows, where the
effective Mach number is small. For example, in the experiment of Gollub's group
\cite{Gollub} the parameter $M$ was about $5\times 10^{-5}$. In this regime the
particles far from the moving boundary are in persistent contact with each other
\cite{Gollub}, inter-particle friction is important (see also \cite{Swinney}),
and the model of inelastic hard spheres (and the Navier-Stokes hydrodynamics) is
inapplicable.

In general, the model of inelastic hard sphere fluid is considered
as a good approximation for dilute and moderately dense flows. Its
validity range for dense flows is not well known \cite{hydrreview}.
Indeed, the particle collision rate increases with the density, so
the assumption of instantaneous collisions, intrinsic in the model
of hard spheres, may become restrictive at high densities. Being
aware of this limitation, we still believe that the model of
inelastic hard spheres can capture some of the physics of dense
flow. Like in many other problems, it is useful to push this model
to an extreme and analyze its predictions (some of which being quite
unexpected as we have shown) in detail. We notice in passing that
the two-phase flow predicted in this work resembles experimentally
observed shear bands: localized regions of ordered granular flow,
coexisting with essentially immobile solid-like regions
\cite{shearbands}.

The future experimental and theoretical work should elucidate the
exact conditions under which shear-induced crystallization develops
in granular flows. Though crystallization under shearing is well
documented in MD simulations \cite{Polashenski,Alam}, a recent
experiment \cite{Behringer} showed that, under certain conditions,
shearing can lead to \textit{disorder}.

Finally, we did not attempt to describe in this work the
shear-induced crystallization \textit{process}. Such a description
is beyond the reach of theory as of present. A promising approach to
this problem should deal, in addition to the hydrodynamic fields,
with an order parameter field and its dynamics. For slow granular
flows such a description is now emerging \cite{Aranson}.

\begin{acknowledgments}
We acknowledge useful discussions with J.P. Gollub,  S. Luding,
P.V. Sasorov, J.-C. Tsai, and A. Vilenkin. The work was supported
by the the Israel Science Foundation (grant No. 107/05) and by the
German-Israel Foundation for Scientific Research and Development
(Grant I-795-166.10/2003).
\end{acknowledgments}

\appendix*

\section{DETAILS OF LINEAR STABILITY ANALYSIS}

Let $u$ and $v$ be the velocity components in the $x$ and $y$
directions, respectively. We assume that the small perturbations
to the uniform steady shear flow do not break the symmetry of the
flow in the $x$ and $z$ directions:
\begin{eqnarray}
&&n(y,t)=n_{0}+n_1(y,t)\,, \nonumber
\\
&&T(y,t)=T_{0}+T_1(y,t)\,, \nonumber
\\
&&u(y,t)=u_{0}(y)+u_1(y,t)\,, \nonumber
\\
&&v(y,t)=v_1(y,t)\,, \label{perturbations}
\end{eqnarray}
where index $0$ denotes the uniform steady shear flow, see
Eqs.~(\ref{simplesolution}), while $T_0 = 1-n_0 =
{\epsilon_\ast}^2$. Linearizing the hydrodynamic equations with
respect to small perturbations, we arrive at
\begin{eqnarray}
&&\frac{\partial n_1}{\partial t}+ n_0\frac{\partial v_1}{\partial
y} = 0\,, \nonumber \\
&&L\,n_0\left(\frac{\partial u_1}{\partial t} +
v_1\,\frac{du_0}{dy}\right)=  \nonumber \\
&&=\frac{\partial}{\partial y}
\left\{\frac{T_0^{1/2}}{1-\lambda-n_0} \left[\frac{\partial
u_1}{\partial
y}+\left(\frac{T_1}{2T_0}+\frac{n_1}{1-\lambda-n_0}\right)
\frac{du_0}{dy}\right]\right\}\,, \nonumber \\
&&L\,n_0\,\frac{\partial v_1}{\partial t} =
-\frac{L}{M}\,\frac{\partial p_1}{\partial y} \nonumber \\
&&+\frac{\partial}{\partial y}\left[
\left(\frac{T_0^{1/2}}{1-n_0}+\frac{4\,T_0^{1/2}}{3(1-\lambda-n_0)}\right)\,\frac{\partial
v_1}{\partial y}\right]\,, \nonumber \\
&&\frac{3}{2}\,L\,n_0\,\frac{\partial T_1}{\partial t} +
L\,\frac{\partial v_1}{\partial y} = \frac{\partial}{\partial
y}\left( \frac{T_0^{1/2}}{1-n_0}\,\frac{\partial T_1}{\partial
y}\right)  \nonumber \\
&&+\frac{M\,T_0^{1/2}}{1-\lambda-n_0}\frac{du_0}{dy}\left[2\frac{\partial
u_1}{\partial y}
+\left(\frac{T_1}{2T_0}+\frac{n_1}{1-\lambda-n_0}\right)\left(\frac{du_0}{dy}\right)\right]
\nonumber \\
&&-\frac{2\,R\,T_0^{3/2}}{1-n_0}\left(\frac{3T_1}{2T_0}+\frac{n_1}{1-n_0}\right)\,,
\label{lineq}
\end{eqnarray}
where $L=(2\,M)^{1/2}\,H/d$, the time is measured in the units of
$H/u_0$, and the value of $H$ is determined by the unperturbed
flow, that is by the uniform steady flow solution. One can check
that the new parameter $L$ (which is absent in the steady state
problem, but enters the linear stability analysis), is fully
determined by the scaled parameters $f$ and $M$ and by the
restitution coefficient $r$.

A further simplification employs time scale separation. Let us
compare the characteristic time scales of the problem. The
acoustic time scale is $\tau_1 =
H\,\epsilon_\ast\,/(P_0/n_c)^{1/2}$, the heat conduction time
scale is $\tau_2 = \tau_1\,(H/d)$, the energy loss time scale is
$\tau_3 = \tau_1\,[H/(dR)]$, and the viscous time is $\tau_4 =
\tau_1\,(H/d)\,({\epsilon_\ast}^2 - \lambda)/{\epsilon_\ast}^2$.
If the density is not too close to the density at which the shear
viscosity diverges, that is if $(H/d)\,({\epsilon_\ast}^2 -
\lambda)/{\epsilon_\ast}^2\, \gg 1$, one can separate the
different time scales and eliminate the (acoustic-like) fast
modes. This is equivalent to the  assumption that the
perturbations evolve in pressure equilibrium with the surroundings
\cite{MeersonRMP}. Using the condition $\partial p_1/\partial t =
0$ instead of the full momentum equation [the third equation in
Eqs.~(\ref{lineq})], we obtain $T_1 = -n_1$. Then, differentiating
the second equation of Eqs.~(\ref{lineq}) with respect to $y$ and
substituting $\partial v_1/\partial y$ from the first equation of
Eqs.~(\ref{lineq}), we arrive at
\begin{eqnarray}
&&L\,\left(1 - {\epsilon_\ast}^2\right)\,\frac{\partial}{\partial
t}\left(\frac{\partial u_1}{\partial y} - \frac{n_1}{1 -
{\epsilon_\ast}^2}\right) = \nonumber \\ &&
\frac{\partial^2}{\partial y^2}
\left\{\frac{\epsilon_\ast}{{\epsilon_\ast}^2 - \lambda}
\left[\frac{\partial u_1}{\partial
y}+\left(-\frac{n_1}{2\,{\epsilon_\ast}^2}+\frac{n_1}{{\epsilon_\ast}^2
- \lambda}\right)\right]\right\}\,, \nonumber \\ &&
-L\,\frac{\partial n_1}{\partial t} \left[ \frac{3(1 -
{\epsilon_\ast}^2)}{2}+\frac{1}{1-{\epsilon_\ast}^2}\right] =
-\frac{\partial}{\partial y}\left(
\frac{1}{{\epsilon_\ast}^2}\,\frac{\partial n_1}{\partial
y}\right) + \nonumber \\ &&
\frac{M\,\epsilon_\ast}{{\epsilon_\ast}^2 -
\lambda}\,\left[2\frac{\partial u_1}{\partial y}
+\left(-\frac{n_1}{2\,{\epsilon_\ast}^2}+\frac{n_1}{{\epsilon_\ast}^2
- \lambda}\right)\right]+
\frac{R\,n_1}{\epsilon_\ast}=0\,,\nonumber
\end{eqnarray}
where we substituted $du_0/dy = 1$ and $T_0 = 1-n_0 =
{\epsilon_\ast}^2$. Consider a single Fourier mode of the
perturbation:
\begin{eqnarray}
&&n_1(y,t)=\tilde{n}\exp(\Gamma t +i k y)\,, \nonumber
\\
&&u_1(y,t)=\tilde{u}\exp(\Gamma t +i k y)\,, \nonumber
\end{eqnarray}
where $k$ is the wave number.  Looking for nontrivial solutions,
we obtain
\begin{equation}
\mbox{det}(\mathbf{A}) = 0\,, \label{det}
\end{equation}
where the elements of the matrix $\mathbf{A}$ are
\begin{eqnarray}
&&A_{11}=-\Gamma\,L + \frac{4\,R^2\,\epsilon_\ast\,k^2}{M^2} -
\frac{R\,k^2}{M\,\epsilon_\ast}\,, \nonumber
\\
&&A_{12}=\frac{\Gamma\,L\,k\,f}{R^{1/2}} +
\frac{2\,R\,\epsilon_\ast\,k^3}{M}\,, \nonumber
\\
&&A_{21}=\Gamma\,L\,\left(\frac{3\,f}{2\,R^{1/2}}+\frac{R^{1/2}}{f}\right)
+ \frac{k^2}{\epsilon_\ast} + \frac{4\,R^2\,\epsilon_\ast}{M}\,,
\nonumber
\\
&&A_{22} = 4\,R\,k\,\epsilon_\ast\,. \nonumber
\end{eqnarray}
Equation~(\ref{det})  yields a quadratic equation for $\Gamma$:
\begin{equation}
a_2\,\Gamma^2 + a_1\,\Gamma + a_0 = 0\,, \label{quadr}
\end{equation}
where the coefficients are
\begin{eqnarray}
&&a_{2}=L\,k^2\,\left(1+\frac{3\,f^2}{2\,R}\right)\,, \nonumber
\\
&&a_{1}=4\,R\,L\,k\,\epsilon_\ast\,\left(1+\frac{f\,R^{1/2}}{M}\right)
\nonumber
\\
&&+\frac{L\,f\,k^3}{R^{1/2}\,\epsilon_\ast}\left(1+\frac{3\,R\,{\epsilon_\ast}^2}
{M}+\frac{2\,R^2\,{\epsilon_\ast}^2}{M\,f^2}\right)\,,\nonumber
\\
&&a_{0}=\frac{2\,R\,k^3}{M}\left(k^2 -
\frac{4\,R^2\,\lambda}{M}\right)\,. \nonumber
\end{eqnarray}

\end{document}